\documentclass[final]{cvpr}

\usepackage{times}
\usepackage{epsfig}
\usepackage{graphicx}
\usepackage{amsmath}
\usepackage{amssymb}

\usepackage{booktabs}
\usepackage{multirow}
\usepackage[usenames,dvipsnames]{color}
\usepackage{colortbl}

\usepackage{pifont}
\usepackage{enumitem}

\definecolor{mygray}{gray}{.9}
\definecolor{mypink}{rgb}{.99,.91,.95}
\definecolor{mycyan}{cmyk}{.3,0,0,0}
\newcolumntype{g}{>{\columncolor{mygray}}c}
\newcolumntype{w}{>{\columncolor{mygray}}l}

\usepackage[pagebackref=true,breaklinks=true,colorlinks,bookmarks=false]{hyperref}

\def\cvprPaperID{8686} 
\def\confYear{CVPR 2021}

\begin{document}

\title{Tuning IR-cut Filter for Illumination-aware Spectral Reconstruction from RGB}
\author{Bo Sun$^1$ \quad Junchi Yan$^2$ \quad Xiao Zhou$^{3}$ \quad Yinqiang Zheng$^4$ \\
$^1$University of Southern California, USA\quad\quad$^2$Shanghai Jiao Tong University, China\\
$^3$Anhui Normal University, China\quad\quad$^4$The University of Tokyo, Japan\\
}


\maketitle
\pagestyle{empty}
\thispagestyle{empty}

\begin{abstract}
\vspace{-4mm}
\ \ \ \ To reconstruct spectral signals from multi-channel observations, in particular trichromatic RGBs, has recently emerged as a promising alternative to traditional scanning-based spectral imager. It has been proven that the reconstruction accuracy relies heavily on the spectral response of the RGB camera in use. To improve accuracy,  data-driven algorithms have been proposed to retrieve the best response curves of existing RGB cameras, or even to design brand new three-channel response curves. Instead, this paper explores the filter-array based color imaging mechanism of existing RGB cameras, and proposes to design the IR-cut filter properly for improved spectral recovery, which stands out as an in-between solution with better trade-off between reconstruction accuracy and implementation complexity. We further propose a deep learning based spectral reconstruction method, which allows to recover the illumination spectrum as well. Experiment results with both synthetic and real images under daylight illumination have shown the benefits of our IR-cut filter tuning method and our illumination-aware spectral reconstruction method. \end{abstract}

\vspace{-3mm}
\section{Introduction}
\vspace{-2mm}
\begin{figure}[t]
\centering
\includegraphics[width=1\linewidth]{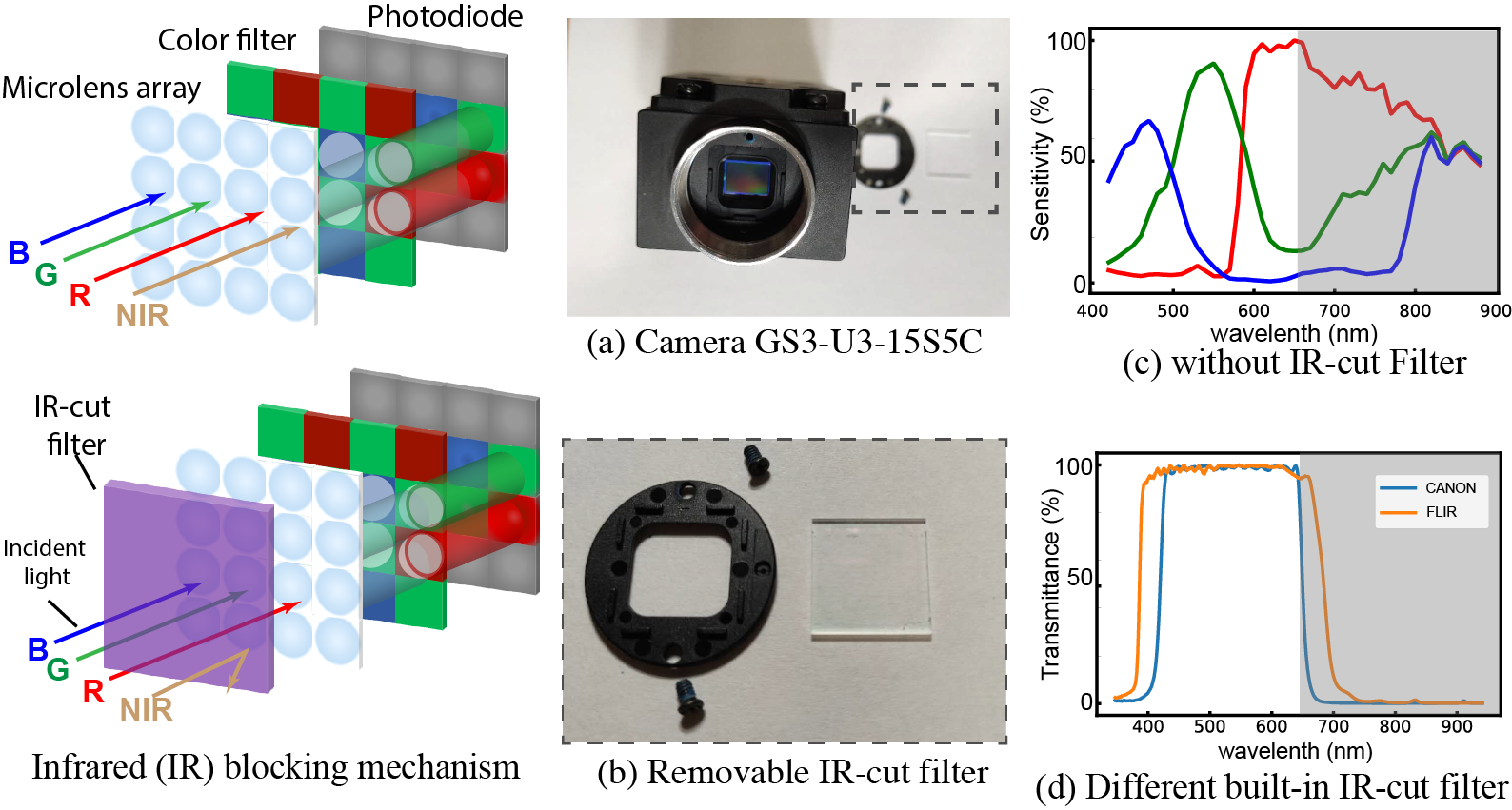}
\caption{Infrared (IR) blocking mechanism of commodity digital color cameras. An IR-cut filter (b) is usually placed in front of the color sensor (a), which can be easily removed. After removing the IR-cut filter, the silicon based sensor can perceive infrared light (c). By measuring the transmittance of the IR-cut filters from different camera makers (d), it is found that the cut-off wavelength can vary a lot, which might affect spectral reconstruction from RGB. Inspired by this mechanism, we propose to design the IR-cut filter in the first place for better and stable spectral upsampling.}
\label{fig:xiangji}
\end{figure}

Hyperspectral imaging (HSI) records detailed spectral information of scene surfaces. It has become an emerging scientific tool in a variety of fields, \eg seed viability estimation~\cite{feng_hyperspectral_2019}, wound healing analysis~\cite{calin_hyperspectral_2015}, non-contact forensic science~\cite{edelman_hyperspectral_2012} and thin-film imaging~\cite{furchner_hyperspectral_2019}. However, hyperspectral acquisition systems are usually scanning based, and remain slow in capture speed \cite{hagen_review_2013,li_fruit_2017}, computational algorithms are therefore in demand to conquer the limitations of conventional hyperspectral imagers.

Recently, methods recovering spectra from a single RGB image have been trend-setting and promising \cite{arad_sparse_2016,arad_filter_2017}, dictionary based approaches and deep convolutional neural networks (CNN) have shown their efficacy in hurdling the non-linear mapping from RGB values to the spectra \cite{nguyen_training-based_2014,shi_hscnn_2018,alvarez-gila_adversarial_2017,zhang_hyperspectral_2019}. It has been proven that the spectral reconstruction accuracy relies heavily on the spectral response of the RGB camera in use \cite{arad_filter_2017}. Inspired by this key observation, data-driven algorithms have been proposed to retrieve the best camera spectral sensitivity (CSS) curves of existing RGB cameras \cite{ferrari_joint_2018}, or even to design brand new three-channel response curves \cite{nie_deeply_2018}.  

However, filter set selection methods \cite{arad_filter_2017,ferrari_joint_2018} can only select from existing CSS databases, and we question the appropriateness of selecting from CSS of commercial cameras designed for human color perception for spectral upsampling. As shown in Fig.\ref{fig:xiangji} (d), some IR-cut filters embedded in commercial RGB cameras tend to cut off a lot of energy beyond 650nm, which will definitely undermine the efforts in~\cite{nguyen_training-based_2014,ferrari_joint_2018,arad_filter_2017} to reconstruct spectra in the 420$\sim$720nm visible range from RGB. In particular, the algorithms have to guess the spectral distribution in the 650$\sim$720nm range on the basis of the observations in the 420$\sim$650nm range, which is obviously error prone. \cite{nie_deeply_2018} extends the search space for CSS curves to the infinite non-negative function space by designing three-channel response from scratch, but it's hardware realization requires a complex co-axis optical system and multiple expensive customized filters in the multi-sensor setup.  


In this paper, we propose to implement a deeply tuned filter to replace the built-in IR-cut filter for better spectral reconstruction. As shown in Fig. \ref{fig:xiangji} (a,b), the IR-cut filter is detached from the color sensor and can be easily removed. As will be conducted in Fig. \ref{fig:hardware}, with our method, one can augment a compatible RGB camera to render spectral signals by simply switching on the customized IR-cut filter, without changing the filter array in front of the silicon sensor. The spectra blocking effect of our designed filter is optimized with end-to-end training, and the invasion to a camera device is minimized. Our solution stands out as a better trade-off between reconstruction accuracy and realization complexity.

\begin{figure*}
\centering
\includegraphics[width=0.97\textwidth]{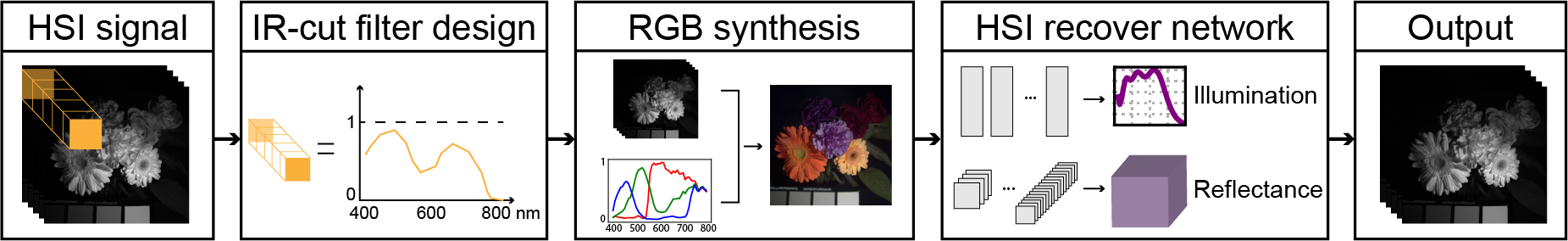}
\vspace{1mm}
\caption{Overview of our end-to-end IR-cut filter design and spectral upsampling network. The IR-cut filter response is simulated as a learnable tensor and will be determined through end-to-end training. There are two separate data flows in our network. The illumination prediction branch ignores spatial relationships and focuses on learning the distribution of illumination spectra. The reflectance reconstruction branch preserves spatial resolution and recovers the scene reflectance. The product of them gives the reconstructed spectra.}
\label{fig:overview}
\vspace{-1.3mm}
\end{figure*}

Another largely untouched aspect of existing RGB-to-spectrum upsampling researches is the physical interaction of illumination and reflectance. By following a simple illumination and reflectance spectral separation model, known as IRSS \cite{zheng_illumination_2015}, we manage to come up with an end-to-end RGB-to-spectrum reconstruction with the ability to estimate incident illumination spectra at the same time. Our design achieves state-of-the-art spectral reconstruction performance and can predict illumination spectra accurately. The overview of our proposed method is shown in Fig. \ref{fig:overview}.

We have conducted extensive experiments on simulated and real spectral images under daylight illumination with different color temperatures, which verify the effectiveness of our IR-cut filter design method, as well as the accuracy and generalization capability of our illumination-aware spectral reconstruction network. To sum up, the major contributions of this work are:
\begin{enumerate}[topsep=-4pt,itemsep=1ex,partopsep=1ex,parsep=1ex]
  \item This work is the first to explore the IR-cut filter response as a tunable factor for spectral upsampling. We also verify the intuition that the optimal cut-off range of the IR-cut filter may not match the objective spectral range. 
  \item We propose a way of addressing the illumination and reflectance separation in the CNN framework, and use it to predict illumination of real outdoor images successfully.
  \item We realize the designed IR-cut filter and verify the effectiveness and plausibility of our proposition with real-life experiments.
\end{enumerate}

\section{Related Work}
\vspace{-2mm}
\subsection{Hyperspectral imager}
\vspace{-3mm}
Traditional HSI systems are usually based on line scanning, which incurs a trade-off between time-consuming operations and spatial resolution. Cameras with mechanically rotating filter wheels or electrically tunable filters are also used to capture spectral images \cite{gupta_hyperspectral_2008,piyawattanametha_tunable_2016}, but it is challenging to capture spectral images of moving objects. Fusion-based approaches present a way of obtaining high resolution spectral images by propagating the spectral information into high resolution RGB images obtained from a hybrid camera system \cite{ma_acquisition_2014,tao_hyperspectral_2020}. Nevertheless, real-time spectral imagers remain prohibitively costly, as precision optics are involved in the system design. 
 
\vspace{-1.mm}
\subsection{RGB-to-spectrum upsampling}
\vspace{-3mm}
To avoid complex design and expensive devices, algorithms have been developed for recovering spectra from RGB values. Injecting proper priors is critical for this under-determined extrapolation problem. Arad \textit{et al.} \cite{arad_sparse_2016} learned a sparse encoding of hyperspectral prior and built a spectrum-to-RGB dictionary via K-SVD. Aeschbacher \textit{et al.} \cite{wu_2017} improved upon \cite{arad_sparse_2016} by incorporating a shallow A+ \cite{tarabalka_segmentation_2010} based re-implementation. Undoubtedly, deep CNN approaches have demonstrated their effectiveness in establishing the nonlinear mapping between RGB values and hyperspectral signals \cite{nguyen_training-based_2014,alvarez-gila_adversarial_2017,can_efficient_2018,li_hyperspectral_2017,han_residual_2018-1}. In particular, Arad and Ben-Shahar \cite{arad_filter_2017} first demonstrated the HSI recovery performance depends heavily on CSS curves, the current state-of-the-art results come from CNN methods that built upon CSS selection \cite{ferrari_joint_2018} and spectral response design from scratch \cite{nie_deeply_2018}. 

\textbf{CSS selection.} Fu \textit{et al.} \cite{ferrari_joint_2018} developed a CSS selection layer with modified network-in-network convolution \cite{lin_network_2014} and retrieved the optimal CSS for spectral reconstruction from existing camera databases \cite{jiang_what_2013}. A drawback is that the search space is severely limited, since the CSS selection layer can only make selection from existing CSS datasets. The information loss incurred in the IR-cut filter embedded in commodity cameras prevents these CSS instances from being optimal for spectral upsampling. 
	
\textbf{Design spectral response from scratch.} Nie \textit{et al.} \cite{nie_deeply_2018} introduced the camera spectral response design into network training via a non-negative 1x1 convolution, and proposed to learn a three-channel spectral responses from scratch. This method extends the search space from existing CSS curves to the infinite non-negative function space. A key limitation of this method comes from the cost and system complexity, since the hardware implementation requires a co-axis optical system and multiple customized color filters.

Physically, the observed CSS is a product of the absolute color sensitivity of camera sensors and the transmittance of the IR-cut filter. Based on this observation, we propose to learn an IR-cut filter response to optimize the camera spectral sensitivity for spectral upsampling without modifying color sensors. In this way, the search space for CSS can be extended to infinite positive function space subject to the constraint that R, G, B channels are complying with the same filtration. The advantage of our method is it brings down the hardware implementation barrier drastically, as we need to implement only one customized IR-cut filter and the invasion into the camera is minimized.

\section{IR-cut Filter Design and Spectral Reconstruction}
\vspace{-3mm}
The key novelty of our work is to augment a consumer-grade camera to a multispectral imager by replacing its IR-cut filter with a deeply learned one. In this section, we present our proposed IR-cut filter design and illumination-aware network architecture. In the next section we provide learning details, datasets and evaluation metrics.
\vspace{-2mm}

\subsection{IR-cut filter spectral response design} 
\begin{figure}[t]
\centering
\includegraphics[width=1\linewidth]{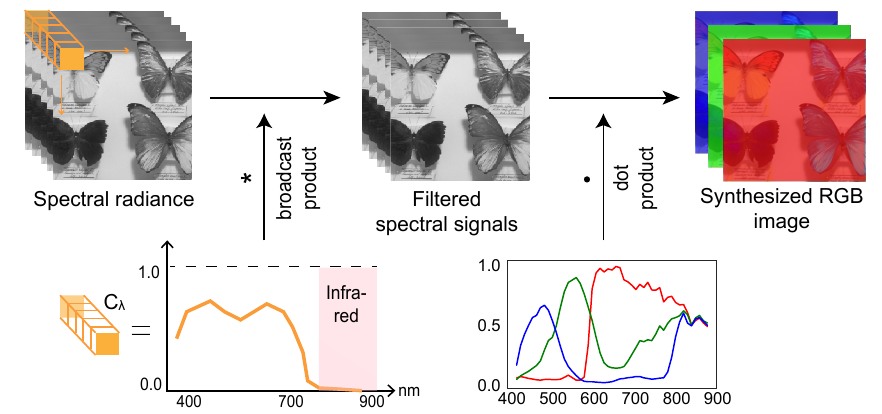}
\caption{Illustration of our IR-cut filter design module. Spectral data cube is first filtered by multiplying $\vec{C_\lambda}$ along spectral axis. RGB images are then synthesized with the filtered spectral signals and camera sensor spectral sensitivity. The synthesized RGB images are later fed to a HSI recovery network.}
\label{fig:design-layer}
\end{figure}

\vspace{-3mm}
The pixel intensity recorded at image cell (x, y) by a linear sensor is an integration over wavelength~$\lambda$ as
\begin{align} \label{eq1}
\mathcal{Y}_c(x, y) = \int_\lambda D(x,y, \lambda)K_c(\lambda) d\lambda,
\end{align}
where $D(x, y, \lambda)$ denotes the spectral energy \textit{arrived} at the sensor and $K_c$ denotes the sensor color sensitivity for $c\in \{R,G,B\}$. For a production camera with IR-cut filter, spectral energy arrived at its sensor equals to the product of incident scene radiance $D_0(x,y, \lambda)$ and light transmittance of IR-cut filter $C_\lambda \in [0,1]$, as
\begin{align} \label{eq2}
\mathcal{Y}_c(x, y) = \int_\lambda D_0(x,y, \lambda)C_\lambda K_c(\lambda)d\lambda.
\end{align}
Assume the number of spectral bands sampled is $M$, in practice equation (\ref{eq2}) is discretized in matrix form as
\begin{align}\label{eq3}
    \mathbf{Y}=(\mathbf{D_0}* \vec{C_{\lambda}} )\mathbf{K_c},
\end{align}
where $\mathbf{Y}\in\mathbb{R}^{N\times 3}$ is the RGB observation, $\mathbf{D_0}\in\mathbb{R}^{N\times M}$ is the scene radiance, $\vec{C_\lambda} \in \mathbb{R}^{1\times M}$ denotes the transmittance of IR-cut filter, and $\mathbf{K_c} \in \mathbb{R}^{M\times 3}$ is the camera color sensitivity. Here $*$ denotes the element-wise product of $\mathbf{D_0}$ and $\vec{C_\lambda}$ along the spectral axis, broadcasting on the spatial axes. This process is illustrated in Fig. \ref{fig:design-layer}.

To recover scene spectra $\mathbf{D}$ from obtained RGB image $\mathbf{Y}$, the neural network needs to approximate a RGB-to-spectra mapping $\mathcal{H}_{\Theta}$ \textit{s.t.}
\begin{align}\label{eq4}
\widehat{\mathbf{D}} =\mathcal{H}_{\Theta}(\mathbf{Y})=\mathcal{H}_{\Theta}((\mathbf{D_0}*\vec{C_{\lambda}})\mathbf{K_c}),
\end{align}
given a training dataset composed of spectral-RGB image pairs $\{(D_i, Y_i)\}_{i=1}^{|N|}$, we register $\vec{C_\lambda}$ as a parameter of the network, and jointly optimize network weights $\Theta$ and IR-cut filter response $\vec{C_\lambda}$ through end-to-end training.

\subsection{IR-cut filter cut-off frequency design}\vspace{-2mm}
Another utility of our IR-cut filter design is to find a data-driven cut-off range for input data in spectra reconstruction problems. Assume the desired spectra reconstruction range is 420$\sim$720nm, all existing methods truncate and use input data exactly of 420$\sim$720nm without justification (even though data of 720$\sim$780nm are often available). But we wonder, because of the continuity of the spectra, can flanking spectral bands outside the objective range provide information useful for reconstructing the objective spectra. 

In our approach, for example, we could let input data be over 420$\sim$780nm and fix $C_\lambda=0$ for $\lambda$ in 730$\sim$800nm to simulate a truncation at 720nm. More interestingly, we can randomly initialize $\vec{C_\lambda}$ between 0 and 1 for all bands over 420$\sim$780nm, and let the back-propagation discover the optimal $\vec{C_\lambda}$. If flanking spectra bands are not useful at all, then upon convergence, the network will set $\vec{C_\lambda}$ to be 0 for all non-informative bands. Otherwise, the network will learn some weights for flanking bands and reach a better solution compared to hard-truncation.

\subsection{Illumination-aware spectral upsampling} \label{aware} \vspace{-2mm}

Illumination and reflectance spectra separation (IRSS) has been a long-standing problem. Mainstream methods assume a subspace model and solve IRSS with low-rank matrix factorization \cite{zheng_illumination_2015,chen_illumination_2017,drew_07}. However, none of the existing CNN-based approaches consider IRSS when reconstructing the spectra. In this study, we integrate into our network the IRSS as a subproblem, and create two deep branches to learn illumination and reflectance separately, the product of which gives the output HSI signal. 

According to the low-rank model of IRSS \cite{zheng_illumination_2015}, the spectral intensity recorded by a hyperspectral imager equals the product of illumination spectra and scene reflectance, in matrix form
\vspace{1mm}
\begin{equation}\label{eq6}
\resizebox{\linewidth}{!}{$
    \underbrace{\left[\begin{array}{ccc}
{d_{11}} & {\cdots} & {d_{1m}} \\
{\cdots} & {\cdots} & {\cdots} \\
{d_{n1}} & {\cdots} & {d_{nm}}
\end{array}\right]}_{\mathbf{D}_{n \times m}} = \underbrace{\left[\begin{array}{ccc}
{r_{11}} & {\cdots} & {r_{1m}} \\
{\cdots} & {\cdots} & {\cdots} \\
{r_{n1}} & {\cdots} & {r_{nm}}
\end{array}\right]}_{\mathbf{R}_{n \times m}} \underbrace{\left[\begin{array}{ccc}
{l_{1}} & {} & {} \\
{} & {\cdots} & {} \\
{} & {} & {l_{m}}
\end{array}\right]}_{\mathbf{L}_{m \times m}} 
$}
\end{equation}
where $m$ is total number of bands sampled and $n$ is the total number of pixels.
Decomposing spectral intensity into the product of illumination and reflectance, equation (\ref{eq2}) now reads
\begin{equation}\label{eq7}
    \mathcal{Y}_c(x, y) = \int_\lambda R(x,y, \lambda)L(\lambda)C_\lambda K_0(\lambda)d\lambda
\end{equation}
and the mapping $\mathcal{H}_{\Theta}$ the network needs to learn becomes
\begin{align}\label{eq8}
\widehat{\mathbf{D}}=
\mathcal{H}_{\Theta}(\mathbf{Y})=
\mathcal{H}_{\Theta}((\mathbf{D_0}*\vec{C_{\lambda}})\mathbf{K_c})\ \dot{=}\ \widehat{\mathbf{R}}*\widehat{\mathbf{L}}\ 
\end{align}
where $\dot{=}$ denotes that the output of the network $\mathcal{H}_{\Theta}(\mathbf{Y})$ are $\widehat{\mathbf{L}}\in\mathbb{R}^{1\times M}$ and $\widehat{\mathbf{R}}\in \mathbb{R}^{N\times M}$, and $*$ denotes channel-wise multiplication broadcasting on the spatial dimensions.

Physically, the illumination spectrum measures the energy distribution of a light source across a wavelength continuum, and it does not depend on the scene surface. Reflectance is the electromagnetic power reflected at an interface that mostly depends on the surface of the material.

Base on the facts above, we design our network structure to learn illumination and reflectance signals separately. RGB signal goes into two deep branches separately. For illumination branch, we stack Squeeze-and-Excitation (SE) block \cite{hu_squeeze-and-excitation_2019} to account for the interdependency and continuity of the illumination spectra, as SE block explicitly models channel-wise dependency. In illumination branch, the spatial resolution decreases while the number of channels goes up in each stacked SE modules. At the end, two transition layers with 1x1 conv bring down the number of channels to $M$, and a global average pooling layer outputs $M$ values as the illumination spectra, where $M$ is the number of spectral bands. For the reflectance branch, we stack some 1x1 convolution layers to learn a starter spectral encoding before stacking dense blocks \cite{huang_densely_2018} for enlarging spatial receptive fields and establishing spatial-spectral relationships. We apply a non-negative 1x1 convolution to bring down the number of feature maps to $M$ because physically optical readings can not be negative. Turns out with this non-negative constraint, the positivity for predicted illumination spectra will be automatically established by the network.

\begin{figure}[h]
\centering
\includegraphics[width=1\linewidth]{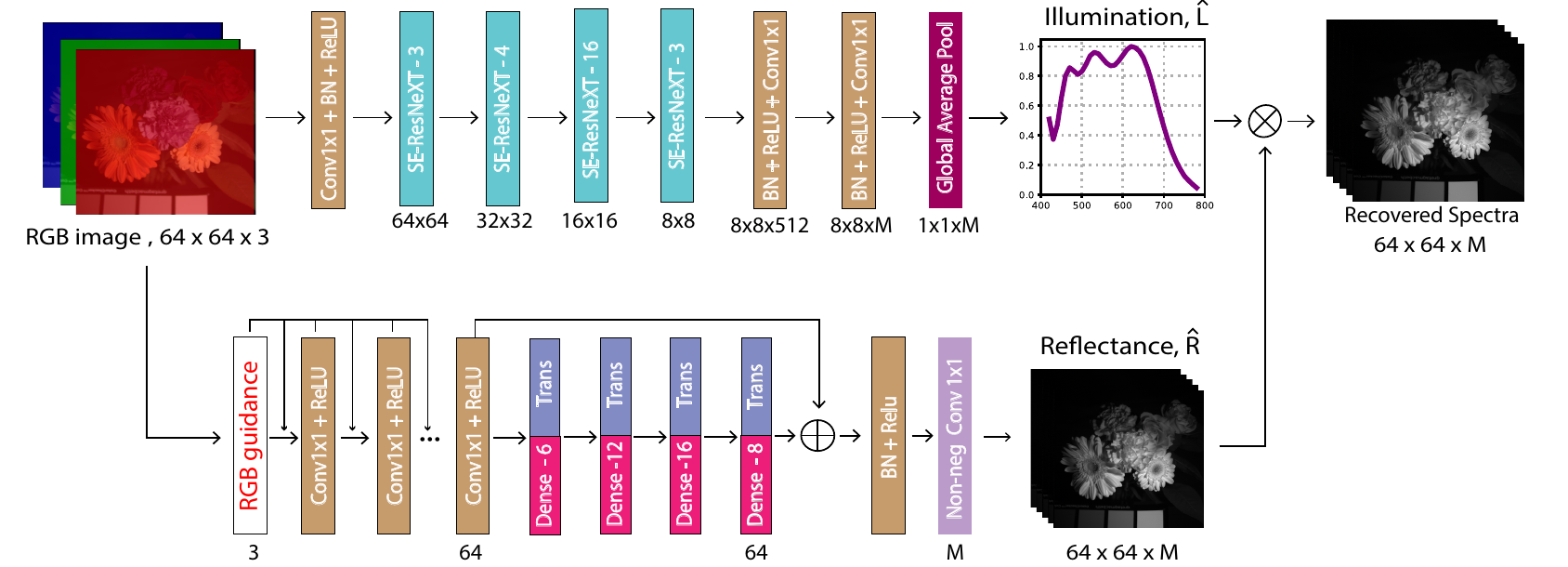}
\caption{Illumination-aware RGB to HSI network. Here shows an example with 64x64 input image patch. The upper is the illumination branch. The lower is the reflectance branch. The spatial resolution does not change in the reflectance branch.}
\label{fig:arche}
\end{figure}

When the ground truth illumination is unknown (hence no supervision signal for illumination), the illumination branch can be turned off, and the reflectance branch will be trained as an end-to-end HSI recovery network. The illustration of the network architecture is shown in Fig. \ref{fig:arche}.

\section{Learning Details} \label{learning-details}
\vspace{-2mm}
\textbf{Implementation.} In our implementation, pre-activation \cite{he_identity_2016} fashion is assured for all residual blocks~\cite{he2016deep} and memory-efficient implementation \cite{pleiss_memory-efficient_2017} of dense blocks is adopted. We add dropout layer at the end of every dense block to counter over-fitting. Regularization is added for all weights via weight decay except for $C_\lambda$, \ie, the IR-cut filter response we intended to design. The network is trained with 128x128 image patches by Adam optimizer \cite{kingma_adam_2017} with a learning rate of $10^{-3}$ and a weight decay of $10^{-4}$. A batch size of 16 is used in our experiments.

\subsection{Training objectives} 
\vspace{-3mm}
The network was trained by minimizing a multi-task loss composed of reconstruction performance, filter smoothness constraint, and an illumination supervision term~(when available) as follows
\begin{equation}
    \mathcal{L} = \mathcal{L}_{MSE} + \mathcal{L}_{Smooth} + \mathcal{L}_{Illu.}
\end{equation}
\textbf{Spectral reconstruction error.} First, the conventionally adopted mean squared error (MSE) between the predicted and ground-truth HSI is optimized,
\begin{equation}\label{eq9}
\mathcal{L}_{MSE} = \frac{1}{N}\sum_{i=1}^{N}||\mathcal{H}_{\Theta}(\mathbf{Y_i}(\mathbf{D_i};\mathbf{K_c}, \vec{C_\lambda})) - \mathbf{D_i}||^2 + \alpha_1||\Theta||^2
\end{equation}

where $\mathbf{D_i}$ is $i$-th input spectral image, and $\mathbf{Y_i}$ is the $i$-th RGB image obtained from our IR-cut design stage, $\Theta$ denotes parameters of network. $\alpha_1=1e^{-4}$ is used for regularization.

\textbf{Smoothness of designed filter.} To facilitate filter manufacturing, the response of the learned IR-cut filter should avoid abrupt changes between neighboring wavelengths. A lag-difference alike loss is added for $C_\lambda $ with small penalty $\alpha_2=1e^{-4}$ to encourage smoothness of the learned filter response and prevent from steep changes between neighboring bands yet avoid collapse to a trivial constant function. 
\begin{equation}
	\mathcal{L}_{Smooth} = \alpha_2\mathcal{L}(\vec{C_\lambda}) = \alpha_2\sum_{i=2}^M(C_i - C_{i-1})^2	
\end{equation}
where $M$ is the number of spectral bands sampled.

\textbf{Illumination guidance}. For synthetic dataset, of which the ground truth illumination is known, we add the supervision signal for illumination spectra recovery to guide the illumination and reflectance separation,
\begin{equation}\label{eq12}
\mathcal{L}_{Illu.} = \alpha_3||I_0(\lambda) - \hat{I}(\lambda)||^2
\end{equation}
where $I_0(\lambda)$ denotes the ground truth illumination and $\hat{I}(\lambda)$ denotes the predicted illumination, with $\alpha_3=0.02$ to balance the scale of losses.

\subsection{Dataset and evaluation metrics}
\vspace{-2.5mm}
We evaluate our filter design layer and spectra reconstruction network with both synthetic and real HSI dataset. For all datasets, 25\% images are held out as test data. The rest images are uniformly incised into 128x128 patches, 85\% patches are used as training data and 15\% patches are used as validation data.

\textbf{Real data}. Dataset TokyoTech \cite{monno_single-sensor_2019} contains 59-band hyperspectral images from 420nm to 1000nm at 10nm increments, presented in form of reflectance spectra. Dataset ICVL \cite{arad_filter_2017} contains 201 hyperspectral images of outdoor scene captured under daylight illumination from 400nm to 1000nm at 1.5nm increments. To be consistent, we sample ICVL data at 10nm increments. The detailed sampling procedure for reducing ICLV data to 10nm increments can be found in the supplementary materials. 

\textbf{Synthetic data}. We synthesized a mixed-illumination dataset called TokyoTech-TC. TokyoTech-TC is synthesized with TokyoTech reflectance and sunlight illumination spectra of color temperature 4000K to 8000K at 1000K increments. In TokyoTech-TC, the ground truth illumination is known, and we use it to evaluate the illumination prediction ability of our network. 

In addition, 34 images from ICVL (objects\_0924-1550 to objects\_0924-1648) contain a white reflector, from which we can estimate the illumination of the scene. These images are used to evaluate the generalizability of our illumination-awareness for outdoor scenes.

\textbf{Evaluation metrics.} 
We evaluate the spectra upsampling performance with three image quality metrics, rooted-mean-squared-error (RMSE), structural similarity index (SSIM \cite{wang_image_2004}) and peak signal-to-noise ratio (PSNR \cite{hore_image_2010}). Smaller RMSE indicates a superior performance, while larger values of SSIM and PSNR are better. The illumination predictions are evaluated by the RMSE and angular error (AE \cite{hordley_re-evaluation_nodate}) between our predicted and the ground truth illumination spectra.

\section{Experiment Results}
\vspace{-3mm}
In this section, firstly we compare our IR-cut filter modules with other filter selection/design methods when fixing to the same HSI recovery network. Secondly, by fixing the filter, we compare the performance of our proposed spectra reconstruction network with current state-of-the-art networks. Then we present our discovery regarding the optimal IR cut-off range for visible spectra ($420 \sim720$nm) reconstruction. Lastly, we demonstrate the illumination prediction power of our proposed method by testing our synthetic data trained network on unseen real data of outdoor scenes.

\vspace{-1mm}
\subsection{Comparison of IR-cut filter designs} 
\vspace{-3mm}
We compare our IR-cut filter design with the CSS selection method \cite{ferrari_joint_2018}, and three-channel spectral responses design \cite{nie_deeply_2018} method. The CSS of Canon 60D with no design is chosen as a baseline. The CSS of the FLIR GS3-U3-15S5C camera without the IR-cut filter (Fig. \ref{fig:xiangji} (c)) is used in our design process. To fairly compare these filter design layers, we use the same HSI recovery network as proposed in \cite{ferrari_joint_2018} for all methods under comparison. Quantitative evaluation results of these filter design layers are shown in Table \ref{table:designcomp}. For convenience, we refer the CSS of Point Grey Grasshopper2 14S5C$\footnote{We use Point Grey Grasshopper2 14S5C--the best CSS selected in~\cite{ferrari_joint_2018}--as the benchmark for CSS selection method.}$ as PG, and the three-channel spectral responses design as RD. 

\begin{table}[h]
\centering
\resizebox{1\linewidth}{!}{\begin{tabular}{c|c|cccc}
\toprule
Dataset           & Metrics & Canon 60D & PG\cite{ferrari_joint_2018} & Ours & RD\cite{nie_deeply_2018} \\
\midrule
\newline{}        & RMSE$\downarrow$  & 4.08 &  3.56   & 3.33     & 3.17   \\
ICVL              & PSNR$\uparrow$  & 35.64 &  38.53  & 39.76    & 39.99  \\
\newline{}        & SSIM$\uparrow$  & 0.974 &  0.980  & 0.985    & 0.987  \\
\hline
\newline{}        & RMSE$\downarrow$  & 4.12 & 3.78    & 3.54     & 3.39   \\
TokyoTech-TC      & PSNR$\uparrow$  & 35.79 & 36.68   & 38.57    & 39.25  \\
\newline{}        & SSIM$\uparrow$  & 0.943 & 0.954   & 0.970    & 0.978  \\
\bottomrule
\end{tabular}}
\vspace{0mm}
\caption{Comparison of filter design layers. Our method outperforms the CSS selection method and slightly underperforms the three-channel design method. This is anticipated since our method is proposed as an in-between solution but excels in the ease of hardware realization.}
\label{table:designcomp}
\end{table}

Our learned IR-cut filter, whose response can be found in Fig. \ref{fig:trunc} (a), achieves a better result than solely selecting from existing CSSs, yet slightly underperforms the three-channel spectral responses design method. This is anticipated as our method essentially optimizes the CSS in infinite non-negative function space as in the freely design method, yet with an extra constraint that RGB curves are subject to the same alteration imposed by the IR-cut filter. In spite of that, our method has a great advantage in its ease in hardware implementation, as will be shown in Section \ref{hardware}.

\begin{figure}[t]
\vspace{-4mm}
\centering
\includegraphics[width=1.5\linewidth]{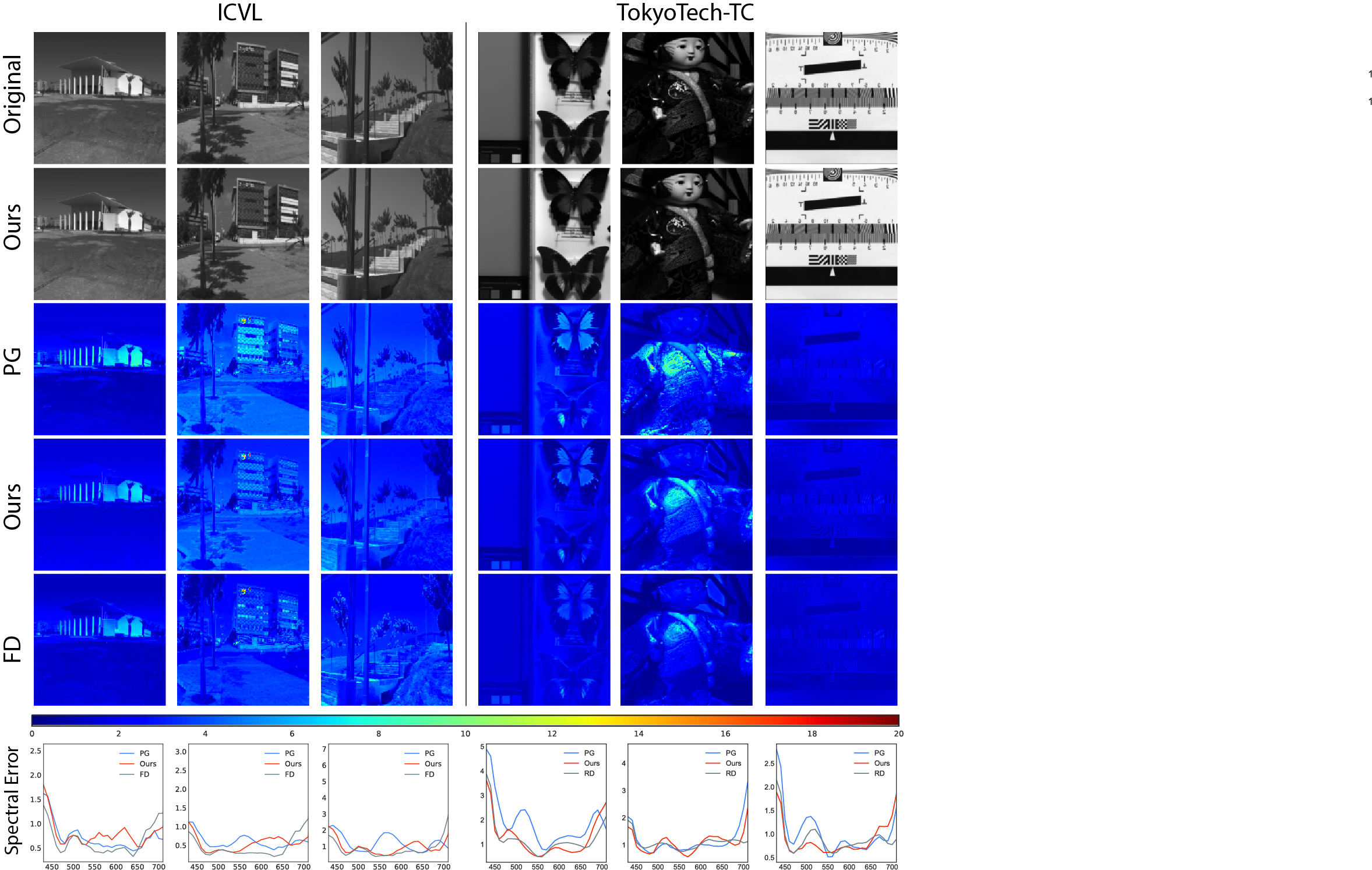}
\vspace{-1mm}
\caption{Visual comparison of spectral reconstruction performance of the three camera response optimization methods. Scenes shown here are randomly selected from ICVL and TokyoTech-TC dataset. By observing the details of the reconstruction and the averaged spectral error, our method is as anticipated an in-between solution compared to CSS selection and three-channel design.}
\label{fig:visual-insp}
\end{figure}

A visual inspection of the performance of different filter design layers on spectra upsampling is shown in Fig. \ref{fig:visual-insp}. The ground truth, our reconstructed result, error images for PG/Ours/RD and RMSE along spectra are shown from top to bottom. Displayed gray image for ground truth and our reconstruction is the 600nm band of the spectra. The error images are obtained as mean absolute deviation (MAD) between the ground truth and the reconstruction across spectra.

\subsection{Evaluation of spectra reconstruction network}
\vspace{-2mm}

Here, we compare our network with the current state-of-the-art HSI reconstruction methods include RBF~\cite{nguyen_training-based_2014}, SR~\cite{arad_sparse_2016} and JCS \cite{ferrari_joint_2018}. To make fare comparison with JCS, we remove filter design layers and use the CSS of Point Grey Grasshopper2 14S5C for both networks. Quantitative evaluation of our implementation of aforementioned methods on different datasets are shown in Table \ref{table:evalcomp}. Results show our network performs consistently better than the neural network based methods RBF and JCS, as well as the sparse representation based method SR. 
\begin{table}[h]
\centering
\resizebox{1\linewidth}{!}{\begin{tabular}{c|c|ccccc}
\toprule
Metrics & Dataset & RBF & SR & JCS & Ours & Ours-IRSS \\
\midrule
\newline{}        & TokyoTech     &  7.32 & 5.49 & 4.32    & \textbf{3.27}    &    - \\
RMSE$\downarrow$              & TokyoTech-TC  &  6.69 & 5.01 & 3.78    & \textbf{3.49}    &   3.54   \\
\newline{}        & ICVL          &  7.71 & 5.20 & 3.56    & \textbf{3.16}    &   3.28   \\
\hline
\newline{}        & TokyoTech     & 0.899 & 0.923 &  0.946    & \textbf{0.979}   &   -  \\
SSIM$\uparrow$              & TokyoTech-TC  & 0.903 & 0.927 &  0.954    & \textbf{0.972}   &  0.962   \\
\newline{}        & ICVL          & 0.919 & 0.935 & 0.980    & \textbf{0.986}   &  0.983   \\
\hline
\newline{}        & TokyoTech     &  28.78 & 31.12 & 34.89    & \textbf{38.97}   &   -   \\
PSNR$\uparrow$              & TokyoTech-TC  &  28.64 & 30.93 & 36.68    & \textbf{38.60}   &  37.48   \\
\newline{}        & ICVL          &  30.13 & 33.42 & 38.53    & \textbf{40.13}   &  39.21   \\
\bottomrule
\end{tabular}}
\vspace{1mm}
\caption{Comparison of reconstruction network on different datasets. The column `Ours' denotes when the illumination prediction is off and the network learns end-to-end RGB-to-HSI mapping, and the column `Ours-IRSS' denotes when the illumination prediction is on and the network needs to learn illumination and reflectance spectra separation as well.}
\label{table:evalcomp}
\end{table}

We note that, when activating the illumination prediction branch in our design, the network needs to solve the illumination and reflectance spectra separation at the same time, therefore the end-to-end measured spectral reconstruction accuracy decreases a tiny bit. In spite of that, our illumination-aware design can estimate the incident illumination spectra to high fidelity, as demonstrated in Section \ref{IA}.

\textbf{Time complexity.} The computation time (in seconds) shown here is benchmarked on 2.4GHz Intel Core i9 CPU and NVIDIA GTEX 2080 Ti GPU. Our methods run as fast as contemporary works that run on GPU and can process about 10 hyperspectral images of size 256x256x31 per second.

\begin{table}[h]\large
\centering
\resizebox{1\linewidth}{!}{\begin{tabular}{cc|c|c|c|c}
\toprule
Image size & SR \cite{arad_sparse_2016} & RBF \cite{nguyen_training-based_2014} & RD (GPU) \cite{nie_deeply_2018} &  Ours (GPU) & JCS \cite{ferrari_joint_2018} (GPU) \\
\midrule
256x256x31 & 2.08s  & 0.20s  & 0.09s  & 0.08s & 0.07s\\
\bottomrule
\end{tabular}}
\vspace{-1.5mm}
\caption{Run time for a single image in seconds (s).}
\label{table:ang}
\end{table}

\subsection{Deeply learned IR-cut filter for visible spectra reconstruction}
\vspace{-2mm}
During our experiments, we have an interesting discovery regarding the cut-off range of input data when training CNN for visible spectral upsampling. When the objective is to reconstruct the $420\sim 720$nm visible spectra, we also include near-infrared bands up to 770nm. Intuitively, if the near-infrared information ($730\sim770$nm) is nothing but noise for visible spectra upsampling, then upon convergence, the network will set the IR-cut filter transmittance of $730\sim 770$nm to be exactly zero, otherwise the algorithm will learn non-zero transmittance for near-infrared bands and reach a better solution with lower RMSE.

\begin{figure}[h]
\centering
\includegraphics[width=1\linewidth]{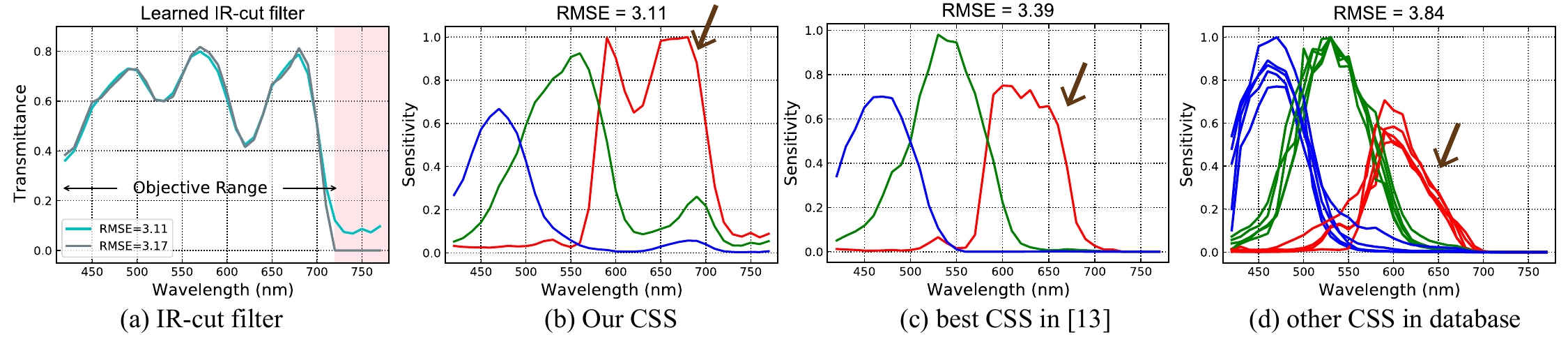}
\caption{Comparison of our obtained CSS and other CSSs in database. (a) Our deeply-learned IR-cut filter, the gray one is forced to have zeros for non-objective wavelength while the cyan one is totally learned through network training. (b) CSS of GS3-U3-15S5C when the cyan IR-cut filter in (a) is applied. (c) CSS of Point Grey, the best CSS in \cite{ferrari_joint_2018}. (d) Examples of other CSSs from database that are selected against in \cite{ferrari_joint_2018}.}
\label{fig:trunc}
\end{figure}

It turns out that letting in near-infrared information can indeed benefit the visible spectral upsampling to some extent. As shown in Fig. \ref{fig:trunc} (a), the deeply learned IR-cut filter has non-zero transmittance for $730\sim770$ nm, in comparison, RMSE increases when the transmittance for $730\sim770$ nm is forced to be zeros. 

Compared with the best CSS from CSS selection method and other CSSs that were selected against in \cite{ferrari_joint_2018}, CSS learned by our method has the highest spectral upsampling fidelity, the spectra reconstruction RMSE are noted in Fig. \ref{fig:trunc}. Remarkably, we notice our learned IR-cut filter has the characteristic of NOT blocking the green and blue channels completely for wavelength over 650nm, and the better HSI recovery capability of the CSS, the later the IR-cut filter starts to block out energies to the infrared end, as can be seen from the trend in Fig. \ref{fig:trunc} (b), (c) and (d). Also, we find the spectral reconstruction error is concentrated on the infrared end in CSS selection method~\cite{arad_filter_2017,ferrari_joint_2018}. We argue this indeed explains the rationale of the selected CSS~--~it is selecting the CSS with the less IR-cut blocking effect. This also proves IR-cut filters in consumer-grade cameras for purpose of blocking out near-infrared light, are suboptimal for HSI reconstruction compared to the learning-based IR-cut filter.


\subsection{Illumination awareness} \label{IA}

\begin{figure}[h]
\centering
\includegraphics[width=1\linewidth]{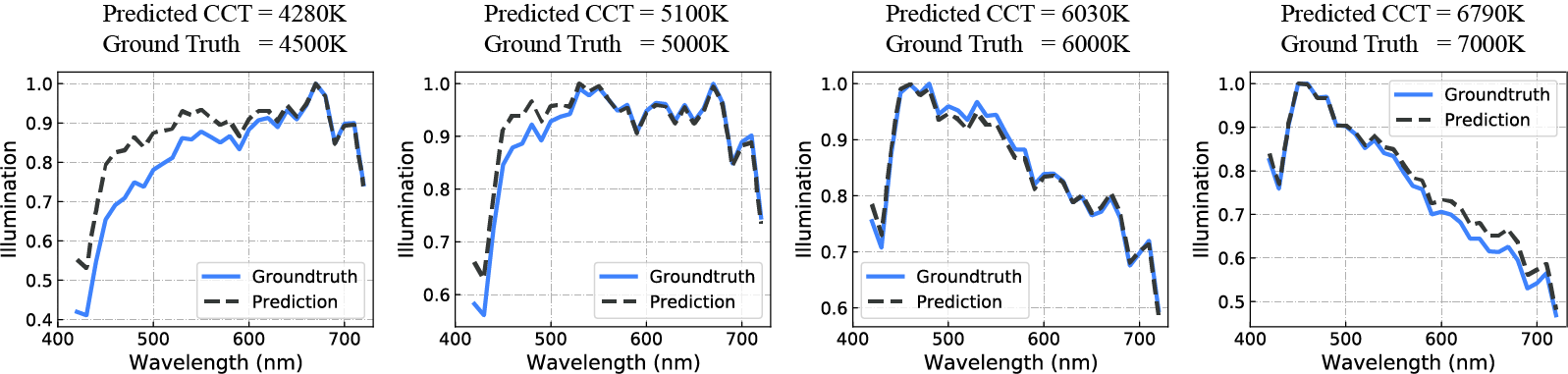}
\caption{Illumination prediction results on TokyoTeck-TC dataset.}
\label{fig:illu-tokyo}
\vspace{-2mm}
\end{figure}

We evaluate the illumination prediction ability of our network in both synthetic and real data. Recall that our TokyoTech-TC dataset is synthesized by reflectance provided by TokyoTech reflectance dataset \cite{monno_single-sensor_2019} and daylight illumination spectra of different color temperature. Fig. \ref{fig:illu-tokyo} shows the predicted illumination and the ground truth when tested on TokyoTech-TC hold-out data. Qualitative results show our method can distinguish and estimate the illumination spectra accurately when trained on such mixed-illumination dataset. Table \ref{table:ang} provides the estimated correlated color temperature (CCT) based on our predicted illumination spectra, and the mean-squared error (MSE) and angular error (AE) between predicted and ground truth illumination, and comparison of our method with well-known white-world assumption \cite{land_lightness_1971}. 

\begin{table}[h]\large
\centering
\resizebox{1\linewidth}{!}{\begin{tabular}{c|c|c|c|c|c}
\toprule
Truth CCT & Est. CCT & MSE Ours$\downarrow$ & MSE~\cite{land_lightness_1971} & AE Ours$\downarrow$ & AE \cite{land_lightness_1971} \\
\midrule
4500K & 4280K  & \textbf{0.0732}   &   0.2242   &   \textbf{0.0648}   &   0.1081   \\
5000K & 5100K  & \textbf{0.0321}   &   0.1198   &   \textbf{0.0320}   &   0.0532   \\
6000K & 6030K  & \textbf{0.0122}   &   0.0900   &   \textbf{0.0137}   &   0.0599   \\
7000K & 6790K  & \textbf{0.0228}   &   0.0878   &   \textbf{0.0212}   &   0.0697   \\
7500K & 7150K  & \textbf{0.0619}   &   0.1000   &   \textbf{0.0570}   &   0.0797   \\
\bottomrule
\end{tabular}}
\vspace{-1mm}
\caption{Evaluation of illumination prediction.}
\label{table:ang}
\end{table}

\begin{figure}[t]
\centering
\includegraphics[width=1\linewidth]{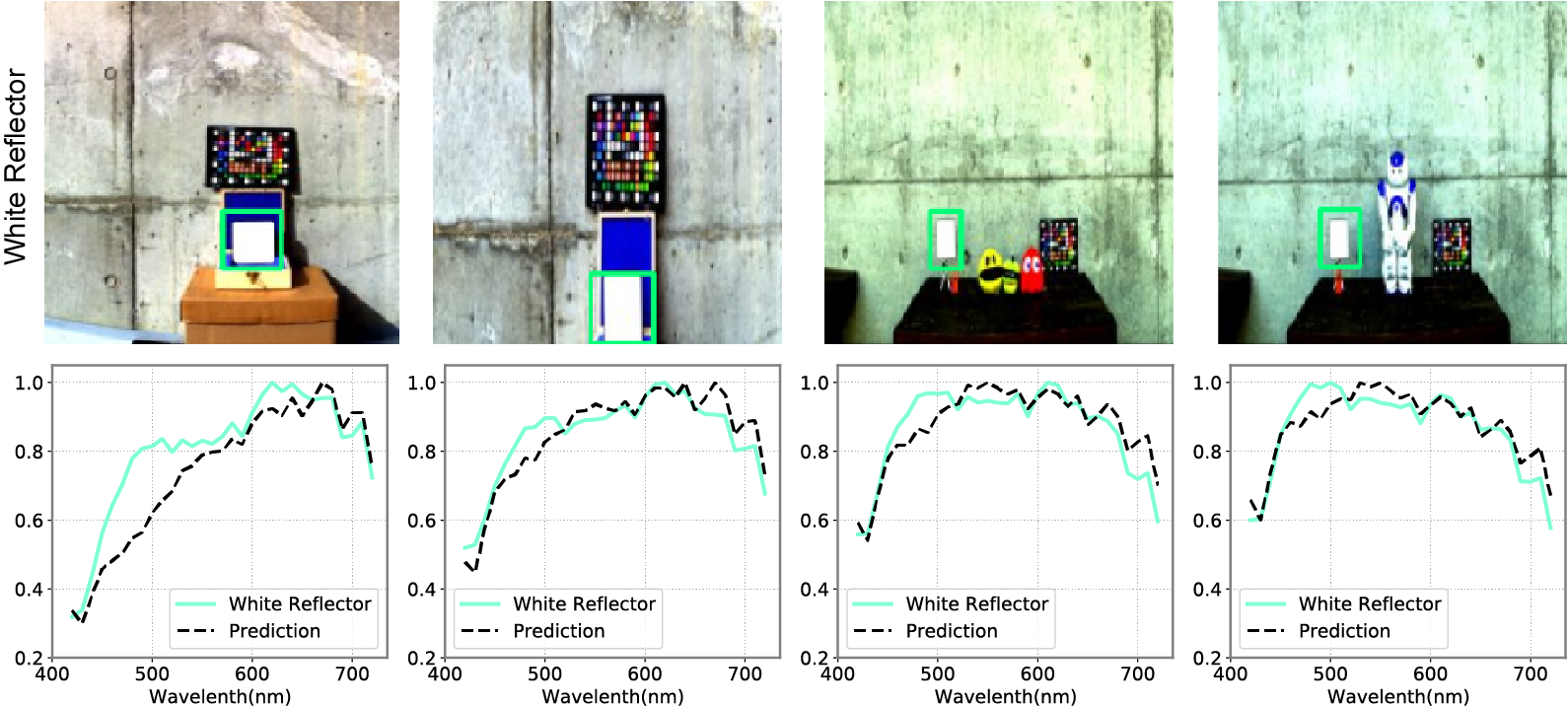}
\caption{Illumination prediction results on ICVL dataset.}
\label{fig:illu-icvl}
\vspace{-4mm}
\end{figure}

Remarkably, our synthesized-data trained model can approximate the illumination spectra quite accurately when tested on ICVL real data
, as shown in Fig. \ref{fig:illu-icvl}. For ICVL, the scenes HSI are captured under outdoor daylight but the ground truth illumination spectrum is not provided. To this end, we select several images with a white reflector, the reflectance spectra of which can be used as an estimation of the illumination.

\vspace{-2mm}
\section{Realization of the Designed IR-cut Filter} \label{hardware}

\begin{figure}[h]
\centering
\includegraphics[width=1\linewidth]{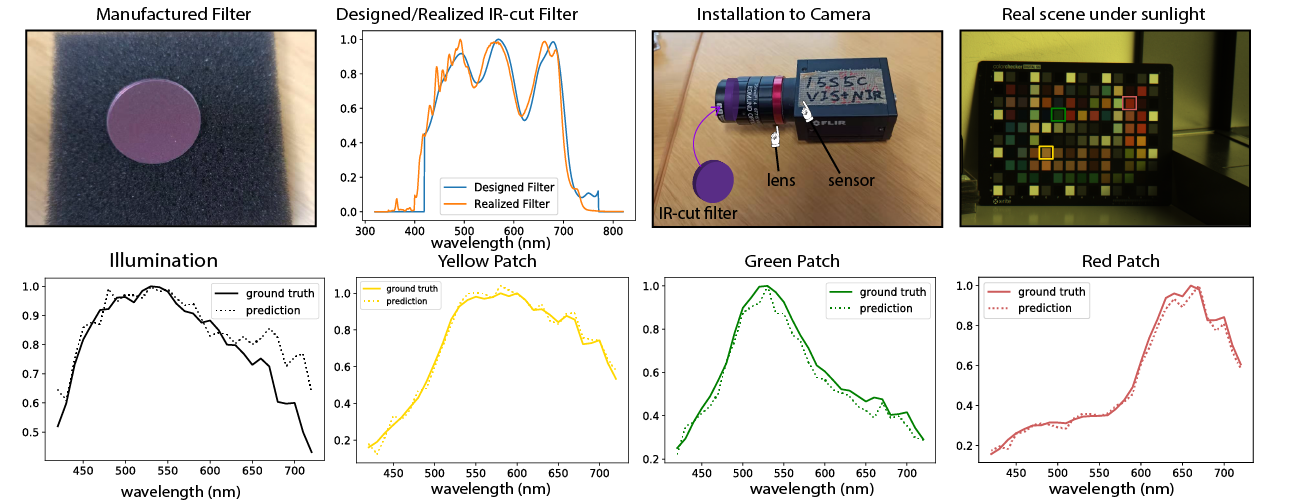}
\caption{Realization of our designed IR-cut filter and spectral reconstruction results of real scenes. The first row shows that the measured response of the realized filter is very close to the design, and the filter can be easily attached to the camera to capture images. The second row shows the reconstruction results for illumination and the reflectance spectra of three checker squares marked in yellow, green and red. Solid line represents spectra measured by our spectrometer and dashed line represents reconstructed spectra. }
\label{fig:hardware}
\end{figure}

As shown in Fig.~\ref{fig:hardware}, we successfully realized our designed IR-cut filter for FLIR GS3-U3-15S5C camera, whose response is very close to the algorithmic design. We install the customized filter to replace the built-in IR-cut filter, without touching the sensor or the circuit. We take pictures with sunlight incident from the window that conforms with the daylight-illumination dataset used to train our network. We compare illumination prediction to the ground truth measured by a spectrometer, and evaluate the reflectance spectra reconstruction performance from pictures taken with our designed filter for different color patches. Results show that, with our realized hardware, the illumination spectrum can be accurately predicted except for a little deviation at the right end of the spectrum. Consistent with simulation results, the reflectance spectra for different color patches can be reconstructed with fidelity.

\begin{figure}[h]
\centering
\includegraphics[width=1\linewidth]{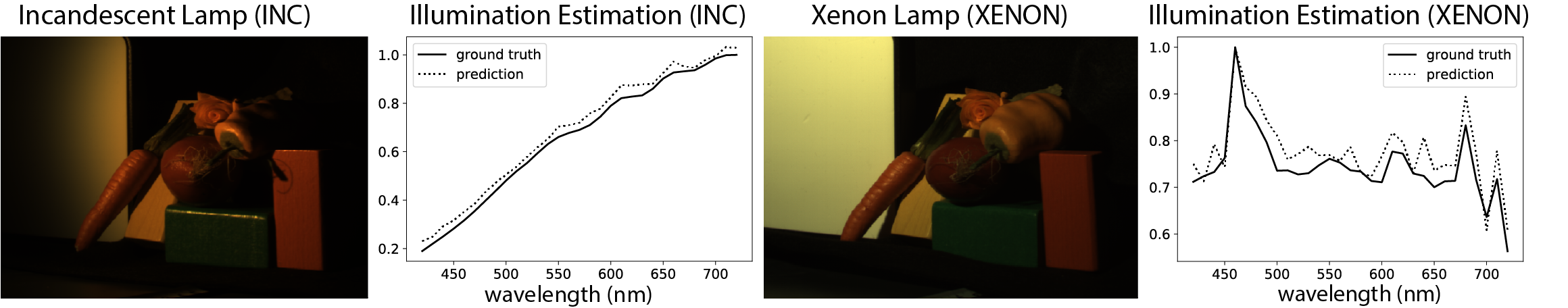}
\caption{Indoor scenes illuminated with incandescent lamp (INC) and Xenon Lamp (XENON), respectively, and their illumination spectra reconstruction results. Solid line represents spectra measured by our spectrometer and dashed line represents reconstructed spectra. }
\label{fig:indoor}
\end{figure}

We also validate the generalizability of our designed filter under different light sources. Fig. \ref{fig:indoor} shows the illumination prediction for incandescent lamp (INC) and Xenon Lamp (XENON). Both light  sources  emit  visible  and near-infrared energies like the sunlight. The results show that our realized filter generalizes to various illumination conditions, and fully demonstrate the plausibility and effectiveness of our proposed methods.

\section{Conclusions}
\vspace{-3mm}
In this paper, we have explored why and how to tune the IR-cut filter for spectral reconstruction from RGB observations. It has been shown that the reconstruction accuracy can be noticeably improved by properly designing the response of the IR-cut filter. We have also incorporated the illumination and reflectance spectra separation model into our newly developed RGB-to-spectrum reconstruction network, which is superior in reconstruction fidelity, and allows to recover the illumination spectrum directly. The feasibility of our idea of tuning IR-cut filter has been further verified by realizing the designed IR-cut filter and using it for accurate spectral reconstruction in real scenarios. As future work, we plan to examine the possibility of accurate spectral reconstruction under general indoor and outdoor illumination. 

\vspace{-4mm}
\section*{Acknowledgement}
\vspace{-4mm}
\small{This work was supported in part by China Major State Research Development Program (2018AAA0100704), NSFC (61972250, U19B2035), Shanghai Municipal Science and Technology Major Project (2021SHZDZX0102), Key Scientific Research Foundation of the Higher Education Institutions of Anhui Province (KJ2017A934), and the JSPS KAKENHI Grant Number 19K20307.}

{\small
\bibliographystyle{unsrt}
\bibliographystyle{ieee_fullname}
\bibliography{egbib}
}

\end{document}